\documentclass{article}
 \usepackage{xcolor}
\usepackage{amsmath}
\usepackage{amsfonts}
\usepackage{amssymb}
\usepackage{graphicx}
\usepackage{stmaryrd}
\usepackage{hyperref}
\usepackage{subfigure}
\usepackage{newpxtext,newpxmath}
\usepackage{lineno}
\newtheorem{Theorem}{Theorem}[section]

\newtheorem {Cor}[Theorem]{Corollary}
\newtheorem {pro}[Theorem]{Proposition}

\newtheorem {Lemma}[Theorem]{Lemma}
\newtheorem {rem}[Theorem]{Remark}
\newtheorem {rems}[Theorem]{Remarks}

\newtheorem {com}[Theorem]{Comment}
\newtheorem {coms}[Theorem]{Comments}

\newtheorem {Definition}[Theorem]{Definition}
\newtheorem {exam}[Theorem]{Example}

  \newcommand{\bcom}{\begin{com} \rm } \newcommand{\ecom}{\end{com}}
\newcommand{\bcoms}{\begin{coms} \rm } \newcommand{\ecoms}{\end{coms}}
\newcommand {\bdef}{\begin{Definition}}
\newcommand {\edefi}{\end{Definition}}
\newcommand {\bl}{\begin{Lemma}}
\newcommand {\el}{\end{Lemma}}
\newcommand {\bethe}{\begin{Theorem}}
\newcommand {\eethe}{\end{Theorem}}
\newcommand {\bp}{\begin{pro}}
\newcommand {\ep}{\end{pro}}
\newcommand {\bcor}{\begin{Cor}}
\newcommand {\ecor}{\end{Cor}}
 \newcommand {\brem }{\begin{rem} \rm }
\newcommand {\erem }{\end{rem}}
 \newcommand {\brems }{\begin{rems} \rm }
\newcommand {\erems }{\end{rems}}

\newcommand {\bcorr}{\begin{corr} \rm }
\newcommand {\ecorr}{\hfill $\lhd$ \end{corr}}
\newcommand {\bex}{\begin{exam} \rm }
\newcommand {\eex}{\end{exam}}
\let\ssection=\section
\renewcommand{\section}{\setcounter{equation}{0}\ssection}
\newcommand {\be}{\begin{equation}}
\newcommand {\ee}{\end{equation}}
\newcommand {\bde}{\begin{displaymath}}
\newcommand {\ede}{\end{displaymath}}
\newcommand {\beq}{\begin{eqnarray*}}
\newcommand {\eeq}{\end{eqnarray*}}
\newcommand {\beqa}{\begin{eqnarray}}
\newcommand {\eeqa}{\end{eqnarray}}

 \def \proof {{\sc{Proof:}}~}

\def \finproof {\hfill $ \square$  \\ }
 \newcommand {\eenu}{\end{enumerate}}

\newcommand {\benu}{\begin{enumerate}}

\def \R{\mathbb R}
\def \E{\mathbb E}
\def \they{y}

\def \P {\mathbb P}

 \def \ind{1\!\!1}

\def \cadlag {{c\`adl\`ag} }

\def \cA { {\cal A}}

 \def \F{{\cal F}}

\def \ff{{\mathbb {F}}}

\def \G{{\cal G}}
\def \gg{{\mathbb {G}}}

\def \hh {{\mathbb{H}}}
 
\def \cG { {\cal G}}

\def \wt  {\widetilde }

   \def\mb{\textcolor{blue} }

    \def\mma{\textcolor{brown}} 
   \makeindex

\usepackage{wrapfig}% To produce the empty space on the left
\usepackage{lipsum}% To produce the empty space on the left
\usepackage{booktabs} % To produce the empty space on the left
\usepackage{xtab} % To produce the empty space on the left
\usepackage{forloop}
\usepackage{etoolbox}

\usepackage{fancyhdr}
\usepackage{fancyhdr}
\author{ Zied Chaieb \thanks{Laboratoire de recherche et d\'eveloppement \mma{QUANTLABS}: the innovative subsidiary of the \mma{QUANTEAM Group}, a consulting firm specializing in Banking and Insurance, in financial services and IT professions, 75008, Paris, France} \and  Djibril Gueye \footnotemark[2] }
\begin{document}

\title{ Pricing zero-coupon CAT bonds using the enlargement of filtration theory: a general framework\footnote{Please address to djibril.gueye{\char'100}quantlabs.fr for correspondence, suggestions and requests for materials.}}

%\title{ Pricing zero-coupon CAT bonds using the enlargement of filtration theory: a general framework}

\maketitle
%\tableofcontents
\begin{center} \includegraphics[scale=0.4]{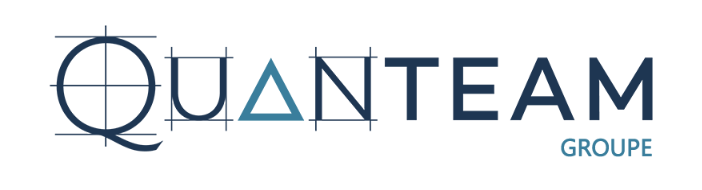}
\end{center}
\begin{abstract}
The main goal of this paper is to use the enlargement of filtration framework for pricing zero-coupon CAT bonds.  For this purpose, we develop two models where the trigger event time is perfectly covered by an increasing sequence of stopping times with respect to a reference filtration. Hence, depending on the nature of these stopping times the trigger event time can be either accessible or totally inaccessible. When some of these stopping times are not predictable, the trigger event time is totally inaccessible, and very nice mathematical computations can be derived. When the stopping times are predictable, the trigger event time is accessible, and this case would be a meaningful choice for Model 1 from a practical point of view since features like seasonality are already captured by some quantities such as the stochastic intensity of the Poisson process. 
We compute the main tools for pricing the zero-coupon CAT bond and show that our constructions are more general than some existing models in the literature. We obtain some closed-form prices of zero-coupon CAT bonds in Model 2 so we give a numerical illustrative example for this latter.

\end{abstract}

 \section{Introduction}
 Catastrophe (CAT) bonds are an alternative to traditional reinsurance intended to hedge against the risks incurred following a natural disaster such as earthquakes,  pandemics, etc. The operation process is as follows.  
 When an insurer or reinsurer wishes to cover a risk in a catastrophic geographical area, they are taken out by investors in the financial markets who perceive so many coupon exchanges. The interest of the issuer is to eliminate the risk of default and increase the available capacity. The operating principle is similar to that of conventional bonds. But in the case of the occurrence of triggering events, the investor loses all or part of the coupons or even the principal of the bond. In the absence of events after a given period, generally between 3 and 5 years, investors recover their initial stakes plus interests. \\
The first catastrophic bonds emerged in 1994, two years after Hurricane Andrew hit, which cost the US \$20 billion. Insurers and reinsurers then realized that natural risks were going to be more and more expensive. CAT bonds were then created as an alternative and complementary to cover extreme risks.\\
In a nutshell, the issuing of catastrophe (CAT) bonds has been and continuous to be essential for insurance companies. Recently, we are witnessing the occurrence of the Coronavirus disease COVID-19 pandemic that causes significant losses and the use of these products would be essential to cover against such losses.
\\

The modeling of CAT bonds is somehow similar to that of credit risk. Indeed, most of the models interested in the construction of the investor's trigger event time (by misuse of language, we may sometimes call it  simply default time)  $\tau$ are based on the structural approach used to price credit derivatives where $\tau$ is the first moment that the aggregate losses process $L$ exceeds the threshold's fixed value $D$ (we  simply call these models first hitting time models). Thus the investor loses all or part of his principal as soon as $\tau$ occurs before the maturity of the CAT bond. The advantage of this type of modeling is that the default time can coincide with the arrival times of catastrophic events. However, even if the aggregate losses approach the threshold (for example $L_t/D=0.9999$, for all $t\geq 0$) as long as the overshoot does not occur, the default mechanism cannot exist. This is not advantageous for the issuer of the CAT (for example a reinsurance company).\\

Several good examples using this approach include the model of Burnecki and Kukla \cite{burnecki2003pricing} who used a  compound doubly stochastic Poisson loss process for aggregate losses where the catastrophe event times are jump times of the standard Poisson process  (i.e., with a constant intensity). Their approach has been used in \cite{hardle2010calibrating} for calibrating CAT bonds for Mexican earthquakes. In \cite{hardle2010calibrating}, by investigating different loss models (such as Pareto, Burr, and Gamma distributions), authors showed that there is no significant impact of these models on the zero-coupon CAT bond prices. The first hitting time model in \cite{burnecki2003pricing} has been extended in  
\cite{ma2017pricing} where the authors used a deterministic intensity rate of the Poisson process in the double Poisson loss model and provide an explicit intensity of the trigger event time as well as a semi-analytical solution for evaluating zero-coupon CAT bonds.\\

In the first hitting time models, default occurs when the aggregate claims process exceeds a specified level (called threshold value). However, in that framework even if the trigger event time can coincide with the arrival time of catastrophe events this interesting property is not highlighted in most of the papers.\\

Another approach to model the investor's trigger event mechanism is to evaluate all the losses until maturity, so the investor loses his principal as soon as the latter exceeds the value of the threshold fixed in advance, otherwise, he recovers its principal plus the coupons. But in this last setting, valuation and payment can be made by taking into account only what happens after maturity. While the default mechanism could even appear just a few months or years after the signing of the contract and for an investor it would be necessary to find out about his default as soon as possible than to wait until the end of the contract.
Among others, it should be referred to Schmidt \cite{Schmidt2014catastrophe}  who used the Shot-Noise process for modeling the aggregated losses process. Shao et al. \cite{shao2016pricing} developed a pricing methodology using a stochastic interest rate framework together with an important focus on two aggregate loss processes such as a compound inhomogeneous Poisson perturbed by diffusion and a general Semi-Markov process.
Mistry and Lombardi \cite{mistry2022pricing} proposed to improve the catastrophe loss estimation by adding  a high spatial resolution for hazard and exposure models. 
However, these approaches fail to take into account the payment at hit (i.e, exactly at the trigger event time, if this occurs before maturity) in the pricing step.\\

These two approaches are equivalent as soon as we are no longer interested in payment at hit and when the percentage of the principal lost by investor in case of the default mechanism is no more stochastic (this is generally the case in CAT bond modeling). Note also that in these two types of approaches, the calibration of the aggregate loss process is essential since the price of CAT bonds is affected by the frequency and also the severity of catastrophic losses.\\
 In other terms, the CAT bond prices depend on the aggregate claims distribution. However in most of the cases, closed form of that  distribution does not exist and one resorts to numerical approximations. For instance, \cite{ma2013pricing}   
has proposed numerical solutions for the loss distribution, in case of compound Poisson process loss process, to compute the price of catastrophe bonds.\\

To the best of our knowledge, only the paper of Jarrow \cite{jarrow2010simple} has used the reduced form approach in credit risk to model CAT bonds where the trigger event time is supposed to admit a deterministic intensity. Even if the latter was not interested in the construction of the investor's trigger event time, his results show the use of the standard Cox model which is the usual model of the reduced-form approach to credit risk. 
 This way of modeling the time to default allows it to obtain a closed-form zero-coupon CAT bond price, and the calibration of the intensity rate can only be done from observed data of the market prices of CAT bonds as studied in the recent paper \cite{bremaud1978changes} which was able to construct a surface of implied intensity rate as a function of maturity and the probability of first loss following a catastrophic event. However, the modeling does not take into account either the severity or the intensity of the losses following the events. On the other hand, in Jarrow's model, the default time cannot, in any case, coincide with catastrophic event times since the reduction of the compensator is absolutely continuous with respect to the Lebesgue measure in a standard Cox model.\\

In this paper, we introduced two models based on the enlargement of filtration theory for pricing zero-coupon CAT bonds. Our approaches will be focused on the information quantified in the CAT bonds. Indeed, we build two models that inherit some credit risk hybrid ones such as the ones developed in \cite{ gueye2021generalized}, \cite{jiao2018modeling} with some ramifications that may be adapted to the CAT bonds modeling context so that they can take into account the aggregate losses. We show through our study that the enlargement of filtration is a suitable tool for CAT bond pricing. The main tool of our models is the fact that the trigger event time may coincide with a strictly positive probability with stopping times of a reference filtration. As such, since these stopping times can be fixed we may construct an accessible trigger event time. In model 2, we also show that in some cases the prices of CAT bonds admit some negative jumps at the catastrophe event time. \\     
Our two models are general frameworks that may lead to a new perspective on modeling CAT bonds. \\   

The paper is organized as follows. 
We first introduce notation and basic notions in Section 2. In Section 3, we present our two models in a general framework by computing the quantities of interest for the pricing of zero-coupon bonds, as the conditional expectations and the dual predictable projection.  In the first model, we show that in the general case where some of the stopping times in the reference filtration are not predictable, our models cover some existing ones in the literature on CAT bond modeling. We also focus on a most meaningful case from the practical point of view, by which the stopping times in the reference filtration are predictable. In the last section, we establish a case study based on simulations of Model 2 where we deal with a particular aggregate loss process called the Shot-Noise process. We illustrate how the jumps in the (Shot-Noise) aggregate loss process induce the jumps in the CAT bonds pricing and compare our results with the ones obtained using a Compound Poisson process where the prices are always continuous.

\section{ Some well-known facts  about default time and enlargement of filtration}\label{dtime}
In this section we recall, for the ease of the reader, some well-known results and definitions.\\  
We consider a probability space $(\Omega, \mathcal{G}, \P)$ and $\tau$  a positive random time defined on  $(\Omega, \mathcal{G})$.  We introduce the right-continuous increasing default process $A_{t}=\ind_{\{\tau \leq t\}}$ associated with $\tau$ and we denote by $\mathbb{A}=(\cA_{t})_{t\geq0}$ the filtration (completed and right-continuous) generated by $A$.
We recall     that, for any process $X$, one has
$\int _u^{ t}L_{s} dA_{s}= \int_{]u,t]}L_{s} dA_{s}=  L_\tau\ind_{\{u<\tau \leq t\}} $.\\
Let $\hh$  a given filtration on $\Omega$.
There exists a unique $\hh$-optional locally integrable variation process $A^{o,\hh}$, called the $\hh$-dual optional projection of $A$, such that
$$\E\left[\int_{0}^{\infty} L_{s} dA_{s}\right]= \E\left[\int_{0}^{\infty} L_{s} dA_{s}^{o, \hh}\right]$$ for any bounded $\hh$-optional process $X$ such that $\E\left[\int_{0}^{\infty} \vert L_{s}\vert d \vert A\vert_{s}\right]<\infty$.\\
There exists a unique $\hh$-predictable locally integrable variation process $A^{p, \hh}$, called the dual predictable projection of $A$, such that
$$\E\left[\int_{0}^{\infty} L_{s} dA_{s}\right]= \E\left[\int_{0}^{\infty} L_{s} dA_{s}^{p, \hh}\right]$$ for any bounded $\hh$-predictable process $X$ such that $\E\left[\int_{0}^{\infty} \vert L_{s}\vert d \vert A\vert_{s}\right]<\infty$.
we shall sometimes call them $\hh$-dual projections of $\tau$. (see  \cite{he2018semimartingale} for more details on dual projections). \\

If $\tau$ is an $\hh$-stopping time, the compensator of $\tau$ is by definition the unique $\hh$-predictable increasing process $J^{\hh}$ such that $J^{\hh}_{0}=0$ and $A_{t}-J^{\hh}_{t}$ is an $\hh$-martingale (see \cite[p.265]{jeanblanc2009mathematical}). Note that $J^{\hh}_{t}= J^{\hh}_{t\land \tau}$.
This compensator $J^{\hh}$ of $\tau$ is nothing else than $A^{p, \hh}$. This property extends as follows:

 \bl \label{Comp}
For any $\hh$-predictable bounded process $H$, the process
$$H_{\tau} \ind_{\{\tau \leq t\}} - \int_{0}^{t\land \tau} H_{s} dA^{p,\hh}_{s} $$ is an $\hh$-martingale.
\el

\proof  This result follows from the fact that
  $$H_{\tau} \ind_{\{\tau \leq t\}} = \int_{0}^{t} H_{s} dA_{s}= (H\centerdot A)_{t},$$
and, for $H$ being $\hh$-predictable,  the $\hh$-dual predictable projection of $H\centerdot A$ is $H\centerdot A^{p,\hh}$ (see \cite{he2018semimartingale}, p. 148, Theorem 5.23).
\finproof
\bdef
A random time $\tau$ is said to avoid $\hh$-stopping times, if for any finite $\hh$-stopping time $\xi$, one has $\P(\tau=\xi)=0$.
\edefi

 \bdef
Let $\vartheta$ an $\hh$-stopping.
\begin{itemize}
\item We say $\vartheta$  to be $\hh$-predictable if there exists an increasing sequence of $\hh$-stopping times $(\vartheta _{i})_{i\geq 1} $ converging to $\vartheta$ such that $\vartheta_{i}<\vartheta $  on the set $\{\vartheta_i  >0\}$, for all $i$. If $\vartheta$ is $\hh$-predictable, ($\ind_{\{\vartheta \leq t \}}, t\geq 0)$  is a predictable process.
 \item  We say that $\vartheta$ is  accessible if $ {[\! [ \vartheta]\! ]} \subset \cup_i {[\![ \vartheta_i]\! ]}$ where $(\vartheta_i)_{i\geq 1}$  are $\hh$-predictable stopping times, with ${[\! [\vartheta ]\! ]}$ denotes the graph of  $\vartheta$ (i.e  ${[\! [ \vartheta]\! ]}=\{(\omega,t)\,: \vartheta (\omega)=t\}\,$).

\item We say that $\vartheta$ is totally inaccessible if it avoids all $\hh$-predictable stopping times (i.e., $\P(\vartheta= \xi<\infty) = 0$ for any   $\hh$-predictable stopping time $\xi$).
\end{itemize}
\edefi

We  now  {work on a filtered probability space $(\Omega, \cG,\ff,\P)$  on which a random time $\tau$ is defined.}  We denote by $Z$ the Az\'ema supermartingale  (see \cite[Subsection 5.9.4]{jeanblanc2009mathematical}, \cite{azema1972quelques}, \cite{nikeghbali2006essay}) associated with $\tau$, which satisfies  $Z_{t}:=\P(\tau > t \vert \F_{t})$.  {Note that $Z_{t}>0$ on  $\{\tau> t\}$}  and $Z_{t-}>0$ on $\{\tau \geq t\}$ (see \cite[Lemma 2.14]{aksamit2017enlargement}).
Then, $A^{p, \ff}$, the $\ff$-dual predictable projection of $A$, is also the predictable part in the Doob-Meyer decomposition of $Z_{t}:= m_{t}- A^{p,\ff }_{t}$ where $m$ is an $\ff$-martingale (see \cite[subsection 2.2,  {page 33}]{aksamit2017enlargement}).
\\
 
 \bdef
Let $\gg=(\cG_{t})_{ t\geq 0}$ be the progressive enlargement of $\ff$ with $\tau$, i.e., $\gg=\ff\lor \mathbb{A}$, which means that $\cG_{t}=\cap _{\epsilon>0} \cG^{0}_{t+\epsilon}$, with  {$\cG^{0}_{s}= \F_{s}\lor \cA_s$ for every $s\geq 0$ }(see, e.g., \cite{jeulin1978grossissement}, \cite{yor1978grossissement}).\\
 The filtration $\gg$ is the smallest filtration satisfying the usual hypotheses containing $\ff$ and turning out $\tau $ into a stopping time.
 \edefi

 \bdef  \textbf{{ The $\ff$-predictable reduction of the compensator of $\tau$}}\\
The process $\Lambda$ given by    \be \label{IntensityDef} \Lambda_t= \int _0^t \ind_{\{Z_{s-}>0\}}\frac{dA^{p,\ff}_s}{Z_{s-}}\ee is $\ff$-predictable and increasing, and, denoting by $\Lambda^\tau$ the process $\Lambda$ stopped at time $\tau$,
        $$A_t-\Lambda _{t\wedge \tau}=A_t-\Lambda^\tau_t=A_t-\int _0^{t\wedge \tau }\frac{dA^{p,\ff}_s}{Z_{s-}} $$ is a $\gg$-martingale (see \cite[Proposition 2.15]{aksamit2017enlargement}).\\
         The process  $\Lambda^\tau$ is   {the} $\gg$-compensator of the default process $A$ (we shall also say compensator of $\tau$) and we call $\Lambda$ the {$\ff$-predictable reduction of the $\gg$-compensator of $\tau$}. \\
         If $\Lambda$   is absolutely continuous with respect to the Lebesgue measure, i.e., $\Lambda_{t} =\int_{0}^{t} \lambda_{s}ds $, then its derivative $\lambda$, that is a non-negative $\ff$-predictable process, is called the $\ff$-intensity rate.
\edefi
We recall that {the random time} $\tau$ avoids all  $\ff$-stopping times (resp. all $\ff$-predictable stopping times) if and only if $A^{o, \ff}$ (resp.  $A^{p,\ff}$) is continuous (see \cite[Proposition 1.43]{aksamit2017enlargement}).  It can be proved that the jump times of $A^{o, \ff}$ are $\ff$-stopping times not avoided by $\tau$. \\
 
\section{Catastrophe bond modeling}
In a filtered probability space $(\Omega, \cG,\ff,\P)$ covering the market uncertainty, we consider an increasing sequence of $\ff$-stopping times $(\theta_{i})_{i}$, with $\theta_{0}=0$. We consider an increasing \cadlag process $L$ (assumed to be independent of $\ff$) to be the aggregate loss process related to a sequence of catastrophe events such that $L_0=0$, $L_{\infty}=\infty$. \\
Let $\tau$ be the catastrophe default time (known also as trigger event time) of a CAT bond contract. \\

\bdef
A zero-coupon CAT bond with maturity $T$ is a contract that pays a fixed amount (called the principal) $P_{\text{cat}}$ at time $T$ if $\tau$ does not occur before $T$ and a fraction $\delta$ (with $0\leq \delta <1$) of the principal if  $\tau$ occurs before $T$. Its payoff is then given by 

$$\zeta=P_{\text{cat}}\ind_{\{\tau>T\}} + \delta P_{\text{cat}} \ind_{\{\tau\leq T\}}. $$

\edefi

We consider $\gg$ to be the enlarged filtration of $\ff$ with $\tau$. Here $\gg$ represents the extra information about the market. Hence, as one can see, the pricing of the CAT bond should be done in $\gg$. Fortunately, the enlargement of filtration theory offers a way to return the pricing in the reference filtration $\ff$. This requires the knowledge of some characteristics of the trigger event time $\tau$ such as the Az\'ema supermartingale and the dual predictable projection in $\ff$. Furthermore, these quantities depend on the model related to $\tau$. In what follows, we introduce two models for which we give those quantities and discuss the pricing of zero-coupon CAT bonds.

\subsection{Model 1}
We define the trigger event time  $\tau$ as 
\be \label{Def::tau} \tau=\theta_{i}~~~  \text{on} ~~~  \{L_{\theta_{i-1}} \leq D <L_{\theta_{i}}\}, ~   \text{for} ~  i\geq 1 \ee
where $D$ is a (finite) positive fixed amount representing a threshold value of the CAT bond.
We consider the function $\Psi$ related to the law of the aggregate loss process  $L$ as $\Psi(t, D):=\P(L_{t} \leq D)$ and note  that  $\Psi(  {\infty} ,D) =1$ hence,  $\tau$ is  almost surely finite. \\
We will need the  following assumption (we shall see why this hypothesis in the following section):\\ 
\textbf{Hypothesis (\textit{A})}: The law $\Psi$ of $L$ is non-increasing with respect to the time $t$.\\

%We consider $\gg$ to be the enlarged filtration of $\ff$ with $\tau$. Here $\gg$ represents the extra information about the market. Hence, as one can see, the pricing of the CAT bond should be done in $\gg$.  This requires the knowledge of some characteristics of the trigger event time such as the Az\'ema supermartingale and the dual predictable projection in $\ff$. In the next section, we give those quantities.   \\

An interesting feature of our approach is that the $\ff$-stopping times $(\theta_{i})_{i}$ are not avoided by the trigger event time $\tau$ and these times can be fixed a priori so that they coincide with the catastrophe arrival times. 

\subsubsection{The  quantities of interest for pricing}
We start by computing the $\ff$-conditional survival law of the trigger event time $\tau$ which allows deriving the Az\'ema  supermartingale $Z$ associated with $\tau$ and then using the Doob-Meyer decomposition of $Z$ for obtaining the dual predictable projection of $\tau$. All these details related to the computation of the $\ff$-conditional survival law of $\tau$ can be seen in the proof of the following proposition.

 \bp \label{Z}
The  Az\'ema supermartingale $Z$ of the trigger event time $\tau$ is given by
\be  \label{Decrea} Z_{t}:=\P(\tau > t\vert \F_{t}) =1-\sum_{i=1}^{\infty} \ind_{\{\theta_{i} \leq t\}} \left[ \Psi(\theta_{i-1},D) - \Psi(\theta_{i},D) \right], ~\text{for} ~ t \in \R^{+} .\ee

\ep

\brem
Note the importance of the assumption \textbf{(\textit{A})} which implies that $Z$ is non-increasing (this can be seen in \eqref{Decrea}).   
\erem
\proof
For all $t,u\in \R^{+}$, one has 
\begin{align*}
\P(\tau > u\vert \F_{t})
=&\sum_{i=1}^{\infty} \P(\theta_{i} >u,  L_{\theta_{i-1}}\leq D< L_{\theta_{i}} \vert \F_{t}).
\end{align*}
This implies that
\begin{align*}
\P(\tau > u\vert \F_{t})=&\sum_{i=1}^{\infty} \P(\theta_{i} >u,  L_{\theta_{i-1}}\leq D \vert \F_{t}) - \sum_{i=1}^{\infty} \P(\theta_{i} >u,L_{\theta_{i}}\leq D \vert \F_{t})\\
=&\sum_{i=1}^{\infty} \P(\theta_{i} >u,   L_{\theta_{i-1}}\leq D \vert \F_{t}) - \sum_{i=1}^{\infty} \P(\theta_{i-1} >u,   L_{\theta_{i-1}}\leq D \vert \F_{t}) \\
=&\sum_{i=1}^{\infty} \P(\theta_{i} >u\geq \theta_{i-1} L_{\theta_{i-1}}\leq D \vert \F_{t})
\end{align*}
where the second equality is due to the fact that $\theta_{0}=0$.  \\ 
Therefore, by using the tower property one obtains \begingroup
\allowdisplaybreaks
  \begin{align*}
\P(\tau > u\vert \F_{t})=&\sum_{i=1}^{\infty} \E[\ind_{\{\theta_{i} >u\geq \theta_{i-1}\}} \P(L_{\theta_{i-1}}\leq D\vert \F_{\infty}) \vert \F_{t}] \\
=&\sum_{i=1}^{\infty} \E[ \ind_{\{\theta_{i} >u\geq \theta_{i-1} \}} \Psi(\theta_{i-1}, D) \vert \F_{t}]\\
 =& \E\left[\sum_{i=0}^{\infty} \ind_{\{\theta_{i} \leq u\}}\Psi(\theta_{i}, D) - \sum_{i=1}^{\infty}  \ind_{\{\theta_{i}\leq u \}} \Psi(\theta_{i-1},D)   \vert \F_{t}\right].
\end{align*}
\endgroup
Since  $\Psi(\theta_{0}, D)=1$, one has
$$\sum_{i=0}^{\infty} \ind_{\{\theta_{i} \leq u\}} \Psi(\theta_{i},D)= 1+\sum_{i=1}^{\infty} \ind_{\{\theta_{i} \leq u\}} \Psi(\theta_{i},D)$$
Hence, it follows
 \begin{align*}
\P(\tau > u\vert \F_{t})=& 1-\E\left[\sum_{i=1}^{\infty} \ind_{\{\theta_{i} \leq u\}} [\Psi(\theta_{i-1},D)-\Psi(\theta_{i},D)] \big\vert \F_{t}\right].
\end{align*}

If $t\geq u$,  one has
\begin{align*}
\P(\tau > u\vert \F_{t})=& 1- \sum_{i=1}^{\infty} \ind_{\{\theta_{i} \leq u\}} [\Psi(\theta_{i-1}, D) -\Psi(\theta_{i},D)]  
\end{align*}
which is due to the fact that the random variables $\ind_{\{\theta_{i} \leq u \}}$ and $ \ind_{\{\theta_{i} \leq u\}} [ \Psi(\theta_{i-1},D) - \Psi(\theta_{i},D)]$ are $\F_{t}$-measurable. \\

In particular, we have 

$$Z_{t}:=\P(\tau > t\vert \F_{t}) =1-\sum_{i=1}^{\infty} \ind_{\{\theta_{i} \leq t\}} \left[ \Psi(\theta_{i-1},D) - \Psi(\theta_{i},D) \right].$$
We can also easily check that $$ \P(\tau >u \vert \F_{u})= \P(\tau >u \vert \F_{\infty})\,, $$  
which implies the immersion property of the model. \finproof

In what follows, we compute the $\ff$-dual predictable projection of $\tau$ using the Doob-Meyer decomposition of $Z$. This decomposition depends on the nature of the $\ff$-stopping times $(\theta_{i})_{i}$. \\

\textbf{Case where some of the $(\theta_{i})_{i}$ are not predictable:\\}
We denote, for any $i$, by $\Lambda^{i}$ the $\ff$-compensator of $\theta_i$, i.e.,  the $\ff$-predictable increasing process $\Lambda^{i}$, with $\Lambda^{i}_{0}=0$, such that
$(\ind_{\{\theta_{i} \leq t\}}- \Lambda^{i}_{t\land \theta_{i}}, t\geq0) $ is an $\ff$-martingale. If it exists, we denote the intensity rate of $\theta_{i}$ by $\lambda^{i}$ (i.e, $\Lambda^{i}_{t\land \theta_{i}}=\int_{0}^{t\land \theta_{i}} \lambda^{i}_{s}ds $, for all $t\geq 0$). 

\bp
 \label{Comp::dual}  The $\ff$-dual predictable projection $A^{p,\ff}$  of the trigger event time  $\tau$ is given by
\be
A^{p,\ff}_{t}= \sum_{i=1}^{\infty} \int_{0}^{t}  { \left(    \Psi(\theta_{i-1}, D)   - \Psi(s, D)\right)} \lambda^{i}_{s}\ind_{\{\theta_{i-1} \leq s< \theta_{i}\}}ds, ~\text{for} ~ t \in \R^{+}. \ee
\ep

\proof
This proposition can be proved as the same way as Proposition 3.15 of \cite{gueye2021generalized} with $\Gamma=0$ and by letting  $\Psi(t, D)$ play the same role as $e^{-\Psi(t)}$.  The main steps of the proof are based on Lemma \ref{Comp}.
\finproof

\textbf{Case where the $(\theta_{i})_{i}$ are predictable:\\}
In this case, the Az\'ema supermartingale $Z$ is predictable, and one has $A^{p,\ff}_{t}=1- Z_{t}$, for all $t  \in \R^{+}$ (i.e., $m=1$, almost surely). In addition, the trigger event time  covered by the $(\theta_{i})_{i}$ is accessible.\\

In the next subsections, we discuss the pricing of zero-coupon bonds by taking into account the two cases.

\subsubsection{The price of the CAT bond}
For simplicity, w assume a constant  interest rate $r$ and $\P$ being the pricing measure. Hence, the price $V_{t}(T)$ of the zero-coupon CAT bond  with principal $P_{\text{cat}}$ and maturity $T$ at time $t\leq T$ is
$$V_t(T)= e^{rt}\E \left[P_{\text{cat}}e^{-rT}\ind_{\{T< \tau\}}+e^{-r\tau}\delta  P_{\text{cat}}\ind_{\{t< \tau \leq T\}}\vert \G_t \right]\,.$$
According to \cite[Proposition 5.1.1]{bielecki2013credit}, we have 
\be  \label{Gen_Prix} V_t(T) e^{-rt}=\ind_{\{t< \tau\}} P_{\text{cat}} e^{-rT}\frac{1}{Z_t} \E[ Z_T\vert \F_t]+ \ind_{\{t< \tau\}}\delta P_{\text{cat}}\; \frac{1}{Z_t}\E\left[ \int_t^T e^{-ru} dA^{p,\ff}_u \vert \F_t \right],\, \forall 0 \leq t\leq T\,.\ee
By consequence, one has

 \begin{align} V_0(T) =& P_{\text{cat}} e^{-rT}\E[ Z_T]+  \delta P_{\text{cat}}\; \E\left[ \int_0^T e^{-ru} dA^{p,\ff}_u \right] \nonumber\\
 =& P_{\text{cat}} e^{-rT} \left(1-\E \left[ A^{p,\ff}_T \right] \right)+  \delta P_{\text{cat}}\; \E\left[ \int_0^T e^{-ru} dA^{p,\ff}_u \right] \label{def::Last}
 \end{align} 
where we have used the fact that $Z=m-A^{p,\ff}$ and since $A^{p,\ff}_{0}=0$ then $\E[m_{T}]=\E[m_{0}]= \E[Z_{0}]=1$ almost surely.
\subsubsection{Case where some of the $(\theta_{i})_{i}$ are not predictable }

Note that the main quantity we need here for obtaining the price at time 0 of the zero-coupon CAT bond is the dual predictable projection of the trigger event time $\tau$. According to equality \eqref{Comp::dual}, this quantity depends on the law of the aggregate losses, the catastrophe arrival times with their intensity rates, and the threshold level.\\     
By combining \eqref{Comp::dual} and  \eqref{def::Last}, one obtains
  \be  \label{0_CATPrice} 
  V_0(T)= P_{\text{cat}} e^{-rT} \left(1-\sum_{i=1}^{\infty} \int_{0}^{T} \E[Q^{i}(s)]ds \right)+  \delta P_{\text{cat}}\sum_{i=1}^{\infty} \int_0^T e^{-ru}  \E[Q^{i}(u)]du ,
  \ee  
where  $Q^{i}(t):=\left(    \Psi(\theta_{i-1}, D)   - \Psi(t, D)\right) \lambda^{i}_{t}\ind_{\{\theta_{i-1} \leq t< \theta_{i}\}}$, for $i\geq1$.
\paragraph{A particular framework of our model}
Here, we consider a particular case with the two $\ff$-stopping times $\theta_{0}$ and $\theta_{1}$ with $\theta_{0}=0$ (i.e., $i\in \{1, 2\}$).  Then from \eqref{Z} the A\'ema supermartingale $Z$ has the following expression

\begin{eqnarray*} Z_{t} =  \left\{\begin{array}{ll}
 \Psi(\theta_{1},D)& \; \text{if}\; \theta_{1} \leq t     \\
1  & \; \text{if}\; \theta_{1} > t .
\end{array}\right.
\end{eqnarray*}  
From \eqref{Comp::dual} the $\ff$-dual predictable projection $A^{p,\ff}$  of the trigger event time  $\tau$ is given by
$$
A^{p,\ff}_{t}=\int_{0}^{t}  { \left(    1- \Psi(s, D)\right)} \lambda^{1}_{s}\ind_{\{0 \leq s< \theta_{1}\}}ds, ~\text{for} ~ t \in \R^{+}.$$

Therefore,  the price $V_{0}(T)$ of the zero-coupon CAT bond that pays a principal $P_{\text{cat}}$ at $T$ if $\tau $  does not occur before $T$  and zero otherwise (i.e. $\delta=0$), is then given by 

\be  \label{MA}
  V_0(T)= P_{\text{cat}} e^{-rT}\left(1-  \int_{0}^{T}   \E\left[ (    1- \Psi(s, D)) \lambda^{1}_{s}\ind_{\{0 \leq s< \theta_{1}\}} \right] ds  \right)
 \ee 
which is similar to the zero-coupon bond price at time 0 obtained in  \cite{ma2017pricing}. 

\bcom
Note in this case that our model is a more general framework according to some existing CAT bond models in the literature (such as the ones in \cite{baryshnikov2001pricing}, \cite{ma2017pricing}). Indeed, the particular model presented here corresponds to the one where the trigger event time is $\theta_{1}$ and can be modeled by a standard Cox time with an intensity rate $\lambda^{1}$.  It suffices to specify the loss process $X$ and the usual models can be recovered. 
 \ecom

\subsubsection{Case where the $(\theta_{i})_{i}$ are  predictable }
In this case, the main quantity for pricing zero-coupon bonds is the Az\'ema supermartingale $Z$, and  since $Z=1-A^{p,\ff}$, the equality \eqref{Gen_Prix} can be replaced by 
$$  V_t(T) e^{-rt}=\ind_{\{t< \tau\}} P_{\text{cat}} e^{-rT}\frac{1}{Z_t} \E[ Z_T\vert \F_t]- \ind_{\{t< \tau\}}\delta P_{\text{cat}}\; \frac{1}{Z_t}\E\left[ \int_t^T e^{-ru} dZ_u \vert \F_t \right],\,    \forall 0 \leq t\leq T.$$
In the case with zero interest rate, one has $V_0(T)= P_{\text{cat}}  (1-\delta) \E[Z_{T}]$, i.e.,

\be  \label{SimplPrice} V_0(T) = P_{\text{cat}}  (1-\delta) \E\left[ 1-\sum_{i=1}^{\infty} \ind_{\{\theta_{i} \leq T\}} \left( \Psi(\theta_{i-1},D) - \Psi(\theta_{i},D) \right)\right].\ee
%In the particular case where the $(\theta_{i})_{i}$ are deterministic then 
%$$ V_0(T) = P_{\text{cat}} e^{-rT}\left[ 1-\sum_{i=1}^{\infty} \ind_{\{\theta_{i} \leq T\}} \left( \Psi(\theta_{i-1},D) - \Psi(\theta_{i},D) \right)\right].$$

Note in this case that the price depends on the law of the aggregate losses, the catastrophe arrival times, and the threshold level. In the particular case where the $(\theta_{i})_{i}$ are deterministic, the computation is simply based on the computation of the aggregate loss distribution.  Note that, in general, closed-form solutions are not easy to obtain for this distribution but numerical algorithms, such as Monte
Carlo and Fourier transformation can be successfully
used to estimate it. However, the problem of estimating that distribution arises when there is low available data, which is the case in general for CAT bonds.   

\subsection{Model 2}
We recall that given an $\ff$-survival process $Z$ and a uniform random variable $U$ in $[0,1]$ independent from $\ff$, one can construct a random time $\tau$ associated with $Z$ by extended the standard Cox construction as (see, e.g., in \cite{jeanblanc2019characteristics} )
\be \label{Coxtau1} \tau:= \inf{\{t \geq 0:  Z_{t} \leq U \}}.  \ee
In this setting, immersion holds and we have 

$$ \P(\tau>t\vert \F_{t})=   Z_{t} .$$
We consider the survival process of the investor in the CAT bond introduced above to be  defined by \be Z_{t}:= e^{-\frac{L_{t}}{D}}, \; \forall\; t\geq 0  \ee and we model the trigger event time of the investor as in \eqref{Coxtau1}. 
It is clear that this is non-increasing with respect to $t$ and $0\leq Z_{t}\leq 1$, for all $t\geq 0$. Furthermore, $Z$ is non-decreasing with respect to the threshold value $D$. This is a valid survival process in CAT bond modeling.   
Let us note that the definition  \eqref{Coxtau1} is equivalent to define to as 
\be \label{Coxtau} \tau:= \inf{\{t \geq 0:  \frac{L_{t}}{D} >\Theta \}} \ee

where  $\Theta$ is a unit exponential random variable independent of the reference filtration $\ff$. We postulate that the $\ff$-stopping times $(\theta_{i})_{i}$ are the jump times of the process $L$.\\

This model belongs to the so-called Generalized Cox model in credit risk (see \cite{gueye2021generalized}). It suffices to set $K:=\frac{L}{D}$. Hence all the characteristics of the trigger default time $\tau$ can be found following \cite{gueye2021generalized}.  As an example of modeling, instead of using a general form of $L$ we just deal with a particular process known as the Shot-Noise process which is an important process used in CAT bonds. 
\subsubsection{Example of the Shot-Noise process.}
We consider the $\F_{\theta_{i}}$-measurable non-negative random variable $\they_{i}$\footnote{The random variable $(\they_{i})_{i}$ are called the shots and are supposed to be i.i.d in the catastrophe loss modeling.} to be the amount of losses at the catastrophe time $\theta_{i}$ and we denote by $\mu$ the jump measure of the marked point process $(\theta_{i}, \they_{i})$ and $\nu$ its compensator (for simplicity, we suppose $\nu$ to be deterministic\footnote{The assumption for $\nu$ to be deterministic is crucial for the random measure $\mu$ to have independent increments (see Th. 6.2.1 in \cite{jacobsen2006}) and allows to avoid more complications.}.  We define the aggregate loss process as \be \label{SN} L_{t}:=\sum_{i\geq 1} \ind_{\{\theta_{i} \leq t\}} H(t-\theta_{i}, \they_{i})=\int_0^t \int_{\R} H(t-s,x)\mu(ds,dx),\,  {\forall t\geq 0}, \ee
where   $H$ is a function $\R_+\times \R  \rightarrow \R_+$ with \be \label{Gg} H(t,x)= H(0,x) + \int_{0}^{t} h(s,x)ds,    \; \;\; \forall t\geq 0,\; x\in \R, \ee where $h$ is a non-negative Borel function on  $\R_+\times \R $. We assume that \be \label{gsi}\int _0^T \int _{\R} h^2(s,x) \nu (ds,dx) <\infty,\, \forall T < \infty .\ee 
%and that there exists a  non negative function $\varphi$  on $\R_+\times \R$ such that  \be \label{lebemaj}\vert h(s,x )\vert  \leq \varphi(x), \forall  (s,x),\,\;\mbox{ with}\, \int _0^T \int _{\R} \varphi( x) \nu (ds,dx) <\infty,\,\forall T< \infty.\ee
  
%On the other issue
The equality \eqref{Gg} guarantees the fact that $L$ is increasing with respect to the time direction and the relation \eqref{gsi} insures the semi-martingale property of $L$ (see, e.g., Lemma 2 of \cite{Schmidt2014catastrophe}).\\

By  using Proposition 3.15 of \cite{gueye2021generalized}, we obtain 
\be\label{lc} Z_{t}(u):=\P( \tau>u \vert \F_t)= c(u) L_t(u),\quad \mbox{for}\;u \geq t\ee  where,   $c(u)=  \exp\big( \int_0^u \int_\R( e^{-\frac{H(u-s,x)}{D} }-1)\nu(ds,dx)\big) , \forall u\in \R_+$  and   $L_{t}(u)=\exp(-\int_0^t \int _\R \frac{H(u-s,x)}{D}\mu(ds,dx)-\int_0^t \int_\R (e^{-\frac{H(u-s,x)}{D}}-1)\nu(ds,dx))$, for any $u\in \R_+$ which is an $\ff$-martingale. In particular, the survival function of $\tau$ is $\P(\tau>u)=c(u)$ and the Az\'ema supermatingale is $Z_{t}=Z_{t}(t)$.\\

Following the proposition 3.14 of \cite{gueye2021generalized}, we have the $\ff$-dual predictable projection of $\tau$ which is given by 
\be \label{Dproj} A^{p,\ff}_{t}=
  \int_0^t \int_\R   Z_{s-}(e^{-\frac{H(0,x)}{D} }-1)\nu(ds,dx)\ee
and the $\ff$-predictable reduction $\Lambda$ of the compensator of $\tau$ satisfies
 $$d\Lambda_t= \int_\R   (1-e^{-\frac{H(0,x)}{D} })\nu(dt,dx),\,\Lambda _0=0.$$
 Furthermore, if $nu$ is continuous,  the trigger event time $\tau$ admits an intensity rate $\lambda$.
 For instance, if $(\theta_i)_{i}$ are  the jump times of a Poisson process $N$ 
  with deterministic intensity function $\lambda^{N}$  such that   $\nu(dt, dx):=F(dx) \lambda^{N}(t)dt $, where $F$ is a distribution function, then   
 \be \label{RateInt} \lambda_{t}=\lambda^{N}(t)\int_\R   (1-e^{-\frac{H(0,x)}{D} })f(x)dx , \,  \text{for all} \; t\geq 0.\ee
Hence the intensity of the trigger event time depends on the threshold value $D$, the loss severity distribution function $F$, and  the claim arrival intensity $\lambda^{N}.$ Note that the intensity rate of $\tau$ is time-varying if the intensity  of the claim arrival  is time-varying.  Furthermore, this quantity is non-increasing with respect to $D$.

\subsubsection{Price of the zero-coupon CAT bonds under the Shot-Noise model.}
We consider a non-negative stochastic interest rate $r$ adapted to a Brownian filtration $\ff^{W}$ independent of the Shot-Noise process $L$ defined in \eqref{SN} . By denoting $\ff^{L}$ as the filtration generated by the $L$, we define the reference filtration $\ff$ as $\ff=\ff^{W}\lor \ff^{L}$. For simplicity, we consider that the principal $ P_{\text{cat}}$ (that we assume to be equal to 1) is fully lost in case of occurrence of the trigger event, i.e, $\delta=0$.  In this setting, we have the following result.

\bl
The price of the zero-coupon CAT bonds is given by 
\beq 
V_{t}(T) &=\ind_{\{t< \tau\}}\exp\left( \int_{t}^{T} \int_{\R} \left( e^{- \frac{H(T -s, x)}{D}} -  1 \right)\nu(ds,dx) - \frac {1}{D}\int_{0}^{t} \int_{\R} \left[ {H(T-s,x)} -{H(t-s,x)}{} \right] \mu(ds,dx)  \right) {Q}_{t}(T),
\eeq
with ${Q}_{t}(T):= \E\left[e^{-\int_{t}^{T} r_{s}ds}\vert \F^{W}_{t}\right]. $
 
\el

\proof
The price $V_{t}(T)$ of the zero-coupon CAT bond is given by 
$$V_{t}(T)=\E\left[ P_{\text{cat}} \ind_{\{\tau >T\}}e^{-\int_{t}^{T} r_{s}ds} \vert \G_{t}\right].$$
Under the general framework of pricing defaultable zero-coupon bonds of \cite[Proposition 5.1.1]{bielecki2013credit}, we can write 
\begin{align*} V_{t}(T)&= \ind_{\{t< \tau\}}  \frac{1}{Z_{t}} \E \left[Z_{T} e^{-\int_{t}^{T} r_{s}ds}\vert \F_{t}\right]=\ind_{\{t< \tau\}}  \frac{1}{Z_{t}}  \E\left[e^{-\int_{t}^{T} r_{s}ds} \E \left[ Z_{T}\vert \F_{T}^{W}\lor\F_{t}^{L}\right] \vert \F_{t}\right]
\end{align*}  
where we have used the tower property and the fact that fact that $e^{-\int_{t}^{T} r_{s}ds}$ is $\F_{T}^{W}\lor\F_{t}^{L}$-measurable.\\Since $L$ is independent of $\ff^{W}$  (hence $Z$ is independent of $\ff^{W}$ ) and $\F^{L}_{t}\subset \F_{t}$, one obtains 
\begin{align*} V_{t}(T)=\ind_{\{t< \tau\}}  \frac{1}{Z_{t}}  \E\left[e^{-\int_{t}^{T} r_{s}ds} \E \left[ Z_{T}\vert  \F_{t}^{L}\right] \vert \F_{t}\right]&=\ind_{\{t< \tau\}}  \frac{1}{Z_{t}}   \E \left[ Z_{T}\vert  \F_{t}^{L}\right]\E\left[e^{-\int_{t}^{T} r_{s}ds} \vert \F_{t}\right]\\
&=\ind_{\{t< \tau\}}  \frac{1}{e^{-\frac{L_{t}}{D}}}   \E \left[ e^{-\frac{L_{T}}{D}}\vert  \F_{t}^{L}\right]\E\left[e^{-\int_{t}^{T} r_{s}ds} \vert \F^{W}_{t}\right].
\end{align*}
We have

$$\E \left[ e^{-\frac{L_{T}}{D}}\vert  \F_{t}^{L}\right]=\E \left[ e^{-\frac{L_{T}}{D}}\vert  \F_{t}\right]=Z_{t}(T)=c(T)L_{t}(T),$$ where $c(T)=  \exp\big( \int_0^T \int_\R( e^{-\frac{H(T-s,x)}{D} }-1)\nu(ds,dx)\big)$  and  \mb{$L_{t}(T)=\exp(-\int_0^t \int _\R \frac{H(T-s,x)}{D}\mu(ds,dx)-\int_0^t \int_\R (e^{-\frac{H(T-s,x)}{D}}-1)\nu(ds,dx))$}. Hence, the result follows by replacing $L_{t}$ by its value given in \eqref{SN}.
\finproof
\subsubsection{A particular Shot-Noise process.} A very tractable Shot-Noise model called the stochastic discounting model (see \cite{Schmidt2014catastrophe}) is given by 
 \be \label{MarkovShot} L_{t}= \sum_{i\geq 1} \ind_{\{\theta_{i} \leq t\}}\they_{i}e^{-\alpha(\theta_{i} -t)},\ee with $(\they_{i})$ some non-negative random variables and  {$\alpha$} a strictly positive parameter. In this case, the process $L$ is a Markovian Shot-Noise process since $H(T-s, x): = xe^{\alpha(T-s)} = e^{\alpha(T-t) }H(t-s,x)$ (see  \cite{schmidt2017shot}) hence, the price of the CAT bond is given by the  following semi-closed form   

\be  \label{Bprice}
V_{t}( T)= \ind_{\{t< \tau\}} \exp\left( \int_{t}^{T} \int_{\R^{+}} (e^{-\frac{xe^{\alpha (T-s)}}{D}} -1)\nu(ds,dx) -\frac{1}{D}(e^{\alpha(T-t) } -1) L_{t}   \right) {Q}_{t}(T)\;
\ee
where ${Q}_{t}(T):= \E\left[e^{-\int_{t}^{T} r_{s}ds}\vert \F^{W}_{t}\right]. $\\
Here we say semi-closed form because of the quantity $Q(T)$ but closed form can be obtained when the interest rate process is an affine process or a polynomial one.\\ 
For example, in the case where the interest rate follows a Cox-Ingersoll-Ross (CIR) model, i.e., $r$ verifies  $$dr_{t}= \gamma_{r}(\theta -r_{t})dt + \sigma \sqrt r_{t} dW_{t}, \ \ \ \ \ r_{0}= x,$$
 where $\gamma_{r}, \theta$, and $\sigma$ are positive parameters and $W$ a Brownian motion, then  
 $Q_{t}(T)=e^{A_{t}(T) -B_{t}(T)r_{t}}$, where $A$ and $B$ verify 
$A_{t}(T)=2\frac{\gamma_{r} \theta}{\sigma^{2}} \ln\left(\frac{2h\ e^{\frac{1}{2}(\gamma_{r} + h)(T-t)} }{h-\gamma_{r} + e^{h(T-t) } (h+\gamma_{r})}\right) $ \ \ and   \ \
 $B_{t}(T)=\frac{2(e^{h(T-t)} -1) }{h-\gamma_{r} + e^{h(T-t) } (h+\gamma_{r})} $, \ where $h=\sqrt{\gamma_{r}^{2} +2\sigma^{2}}\ .$

\bcom
One of the  specificities of this model is that the arrival catastrophe events induce some jumps in the CAT bond prices at the time of occurrence of those events. Indeed let us consider the $\ff$-adapted process $\wt V(T)$ such that
$V_t(T)\ind_{\{t<\tau\}}=\wt V_t(T)\ind_{\{t<\tau\}}$. This process $\wt V(T)$ always exists and admits some negative jumps with sizes given by   \be  \label{jumpPrice}\Delta \wt V_{\theta_{i}} =\wt V_{\theta_i-}  (e^{-(\frac{H(T-\theta_i,\they_i)}{D}-\frac{H(0,\they_i)}{D})}-1)\,.\ee

This can be easily seen by using the results of example 4.2 of \cite{gueye2021generalized} and by noting that the filtration $\ff^{W}$ that is a Brownian one supports only continuous martingales. The process $\wt V(T)$ is called pre-default price in credit risk modeling. In the same vein, we may call it a pre-trigger price in the CAT bond modeling.  
\ecom

 We now consider the $\ff$-stopping times $(\theta_{i})_{i}$ to be the jumps times of a time-inhomogeneous Poisson process $N$ with intensity function $\lambda^{N}$ and consider the compensator measure $\nu(dt, dx):=F(dx) \lambda^{N}(t)dt $, where $F$ is a distribution function. Therefore, the price of the CAT bond given in \eqref{Bprice} becomes
 \be  \label{BpriceBis}
V_{t}( T)=\ind_{\{t< \tau\}} \exp\left( \int_{t}^{T} \int_{\R^{+}} (e^{-\frac{xe^{\alpha (T-s)}}{D}} -1)F(dx) \lambda^{N}(s) ds -\frac{1}{D}(e^{\alpha(T-t) } -1) L_{t}   \right) {Q}_{t}(T).
\ee
Hence, the price of the zero-coupon CAT bond  is affected by the threshold value $D$,  the claim arrival intensity $\lambda^{N}$, the interest rate uncertainty, the distribution of the shots $F$, and the speed of the growth $\alpha$ of the impulsion function $H$ of the Shot-Noise. These two last parameters constitute the severities of the losses. \\
Furthermore, if $\lambda^{N}$ is constant, we have 
  \be  \label{BpriceBisbis}
V_{t}( T)=\ind_{\{t< \tau\}} \exp\left( \lambda^{N}\int_{t}^{T}\left(\phi(\frac{e^{\alpha (T-s)}}{D}) -1 \right)ds -\frac{1}{D}(e^{\alpha(T-t) } -1) L_{t}   \right) {Q}_{t}(T)
\ee
 where $\phi(u)$ is the Laplace transform of the distribution $F$ evaluated at $u\in \R^{+}$. The Laplace transform is not always explicit but some approximations have been introduced for some distributions. For instance, if $F$ is the log-normal distribution with paramters  $\mu \in \R$  and  $\sigma>0$  then a closed-form of $\phi$ does not exist but numerical approximations can be used to compute $\phi$ (see, e.g, in \cite{asmussen2016laplace}, \cite{ miles2018laplace}).
For instance following  \cite{asmussen2016laplace} we have, by setting $J(s):=\frac{e^{\alpha (T-s)}}{D}$,
$$\phi(J(s))= \frac{e^{- {\mu}{J(s) }}}{\sqrt{1+\mathcal{W}( {\sigma^{2}}{J(s)})}} \exp\left(- \frac{1}{2\sigma^{2}}\mathcal{W}({\sigma^{2}}{J(s)})^{2} -  \frac{1}{\sigma^{2}}\mathcal{W}({\sigma^{2}}{J(s)}) \right)$$ where $\mathcal{W}$ is the Lambert function (see, e.g., \cite{corless1996lambertw}). \\
When $F$ is a Pareto distribution function, i.e.,  in the form 
$$F(x)=1-\frac{b^{a}}{b+x}\;\;  x>0,\;  a>0,\; b>0.$$ Then $\phi $ can be setting using the result of \cite{kotz2006laplace}, i.e., 
  $$\phi(J(s))=a(bJ(s))^{(a-1)/2} \exp\left(bJ(s)/2\right) W_{-(a+1)/2,-a/2}(bJ(s))$$ where $W_{\kappa, \mu}$ is the Whittaker function (see \cite{whittaker1903expression}) .

\bcom 
\begin{itemize}
\item
In the case where $\alpha=0$, the $(\theta_{i})_{i}$ being the jumps times of an homogeneous Poisson process $N$ with intensity $\lambda^{N}$ and the shots $(\gamma_{i})_{i}$ being i.i.d  and independent of $N$, the Shot-Noise process $L$ defined in \eqref{MarkovShot} is a Compound Poisson process.
  \item In this case the CAT bond pre-trigger price is continuous.  
\end{itemize}
\ecom

\section{Illustrative example of the CAT bonds prices in the case with Shot-Noise processes}
In this subsection, we compute the term structure of the pre-trigger price of a zero-coupon CAT bond with principal $P_{\text{cat}}= 1$ and maturity of $3$ years using the Shot-Noise model.  As such, we use a threshold value $D=10000$ and a CIR interest rate model with parameters ($r_{0}=0.0204$, $\theta=0.0204$, $\gamma_{r}=0.0884$, and $\sigma=0.0477$) extracted from \cite{mistry2022pricing} and whose values are calibrated from US treasury yield curve for the period ranging in [1994-2013]. We generate a sample path of the aggregate loss amounts from the Shot-Noise process defined in \eqref{MarkovShot} where the underlying Poisson process is assumed to be homogeneous with parameter $\lambda^{N}=0.5 $. For this purpose, we consider the distribution $F$ of shots to be log-normal with parameters ($\mu=6.387$, $\sigma=0.153$) obtained by \cite{mistry2022pricing} from the calibration of their simulated loss data and a speed of the growth of the catastrophe events  $\alpha=0.8$. Our choice of the log-normal distribution for the shots allows us to make an easy comparison, in terms of the behavior of the CAT bond prices, with a compound Poisson aggregate loss process by only setting $\alpha$ to 0. However, one could choose an exponential distribution with a very small parameter and a small claim arrival rate $\lambda^{N}$ so that we could get closer to the realities of the catastrophe losses. The simulation of the Shot-Noise is done using algorithm 2.1 of \cite{scherer2012shot}.

 \begin{figure}[!ht]
\begin{center}
\includegraphics[width=16cm, height = 6cm]{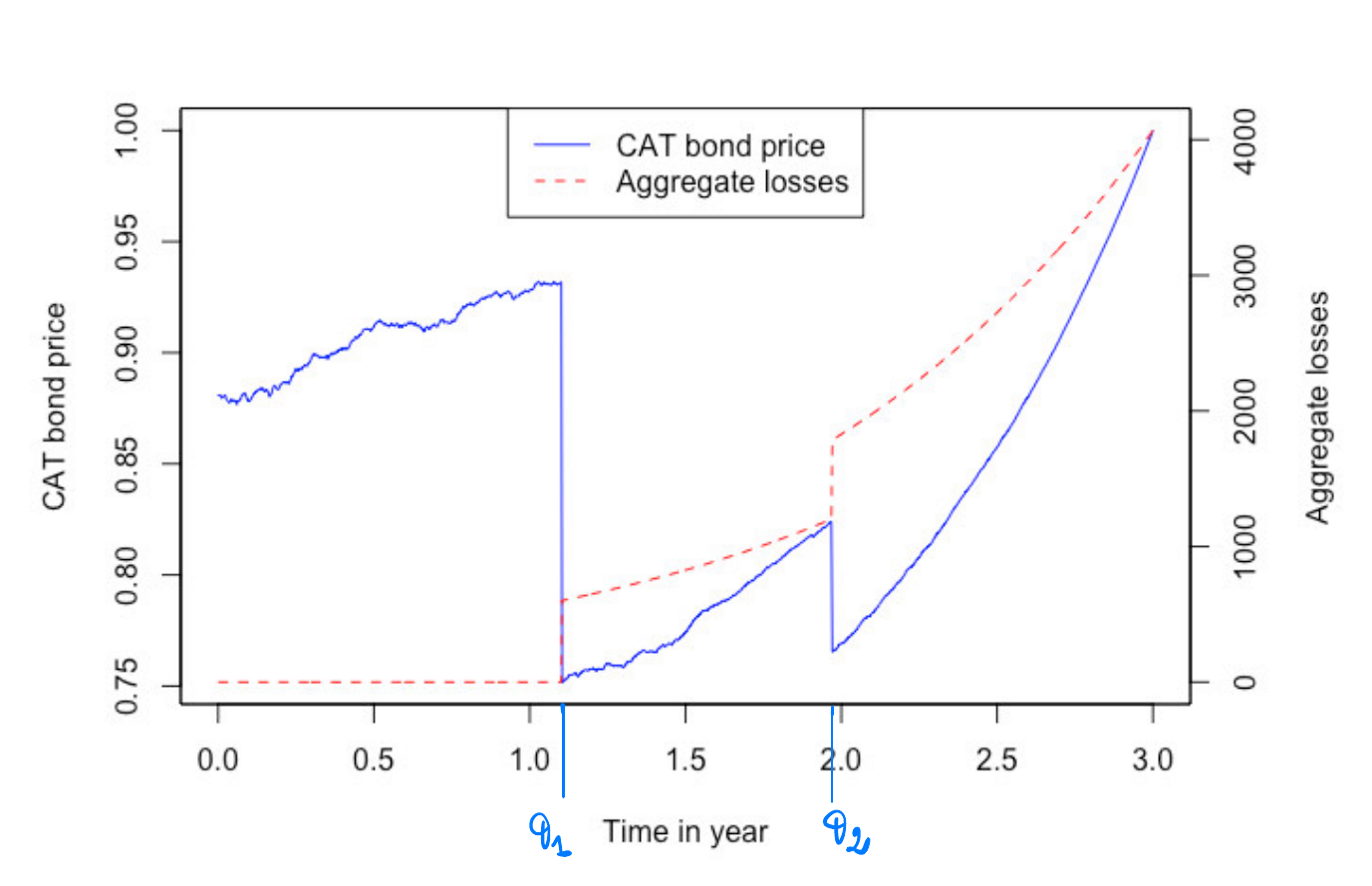}

\end{center}
\caption{CAT bond price vs Aggregate losses}  \label{Fig1}
\end{figure}

In Figure \ref{Fig1}, we represent the term structure of the zero-coupon CAT bond pre-trigger price (the blue curve) as a function of time in years along with the corresponding path of the aggregate losses (in red) generated from the Shot-Noise process $L$.
We note two jumps in aggregate losses at times $ \theta_1=1.104$ and $\theta_2= 1.971$  with jump sizes respectively to equal  $601.8668$ and $582.0399$. This automatically induces negative jumps at the same times, in the price of CAT bond with sizes corresponding respectively to  $0.17989792$ and $ 0.05843795$  computed using the equality \eqref{jumpPrice}.  One can be surprised at the difference in jump sizes during the two events. Indeed, despite the larger size of the second jump of the Shot-Noise, the size of the first price jump remains higher than that of the first. However, in terms of ratio, we can see that it is reasonable since the size of the first price jump represents 19.309152\% of its last value before the jump while that of the second is  7.092837\%  of its last value just before the moment of the jumps. We generally observe an increasing trend for the CAT bond price in between the jump times. For more illustrations, Figure \ref{Sc1} of the appendix shows different scenarios for the behavior of the time-varying CAT bond price according to the trajectory of the aggregate losses.  
 \addtocounter{figure}{1}
 \begin{figure}[!ht]
\begin{center}
\includegraphics[width=16cm, height = 6cm]{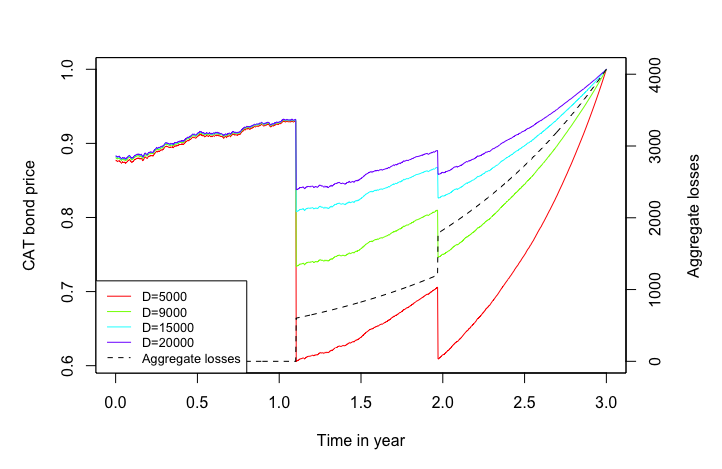}

\end{center}
\caption{CAT bond price with respect to different threshold values}  \label{Fig2}

\end{figure}

Now for a given trajectory of aggregated losses, we fix different threshold levels (5000, 9000, 15000, 20000) and compute for each Threshold the value of the zero-coupon CAT bond at time $0\leq t\leq T$. 
Figure \ref{Fig2} presents the time-varying CAT bond price according to these levels of threshold. Unsurprisingly, the results show that the CAT bond price is non-decreasing with respect to the threshold values. Indeed, Higher threshold value leads to low probability of exceeding it, which leads to the increase in the CAT price. 
This satisfies the results obtained in \cite{ma2017pricing,ma2013pricing} among others. Different scenarios of this analysis can be seen in Figure \ref{Sc2} of the appendix. 
\\

From a standpoint to show the importance of using the Shot-Noise process in our framework, we sample one path of the aggregate losses from a Compound Poisson process (CPP) using a log-normal distribution for the severity distribution and a homogeneous Poisson process with the same parameters as the ones used for the Shot-Noise (i.e., $\mu=6.387$, $\sigma=0.153$ and $\lambda^{N}=0.5 $). We also compute the time-varying pre-trigger price of the CAT bond introduced above using the CPP and the same trajectory of the CIR interest rate.  Figure \ref{FigSNCPP} represents the path of the CAT bond price (black curve) computed using the CPP with the corresponding trajectory of the CPP (green curve) and a trajectory of the CAT bond price obtained from the Shot-Noise (in blue) also along with the corresponding Shot-Noise path (in red).  The results show two jumps of the compound Poisson process at times $\theta_1=0.549$ and $\theta_2=1.674$ with jump sizes respectively equal to  $599.3870$ and $608.2323$. Despite the jumps of the CPP, we observe a continuity everywhere of its corresponding CAT bond price. The last finding shows that it is more convenient to use Shot-Noise processes in our framework than using a Compound Poisson process for the aggregate loss modeling.  

 \addtocounter{figure}{1}
 \begin{figure}[!ht]
\begin{center}
\includegraphics[width=16cm, height = 6cm] {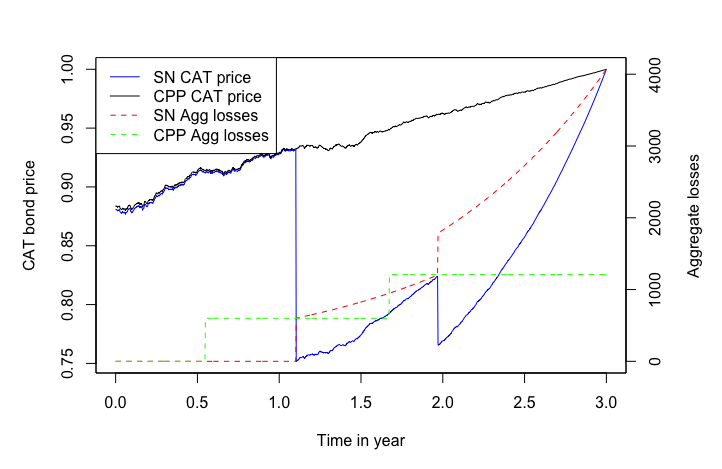}

\end{center}
\caption{CAT bond price vs Aggregate losses}  \label{FigSNCPP}

\end{figure}

In Figure \ref{FigInt}, we plot the surface of the CAT bond price with respect to different maturities and threshold values. Unsurprisingly, as it has been shown by most of papers in CAT bonds modeling (such as for example in  \cite{burnecki2003pricing}, \cite{shao2016pricing}, \cite{mistry2022pricing}),  we observe that the price is non-increasing in the maturity direction and non-decreasing in the threshold value direction.   
\addtocounter{figure}{1}
 \begin{figure}[!ht]
\begin{center}
\includegraphics[width=12cm, height = 6cm] {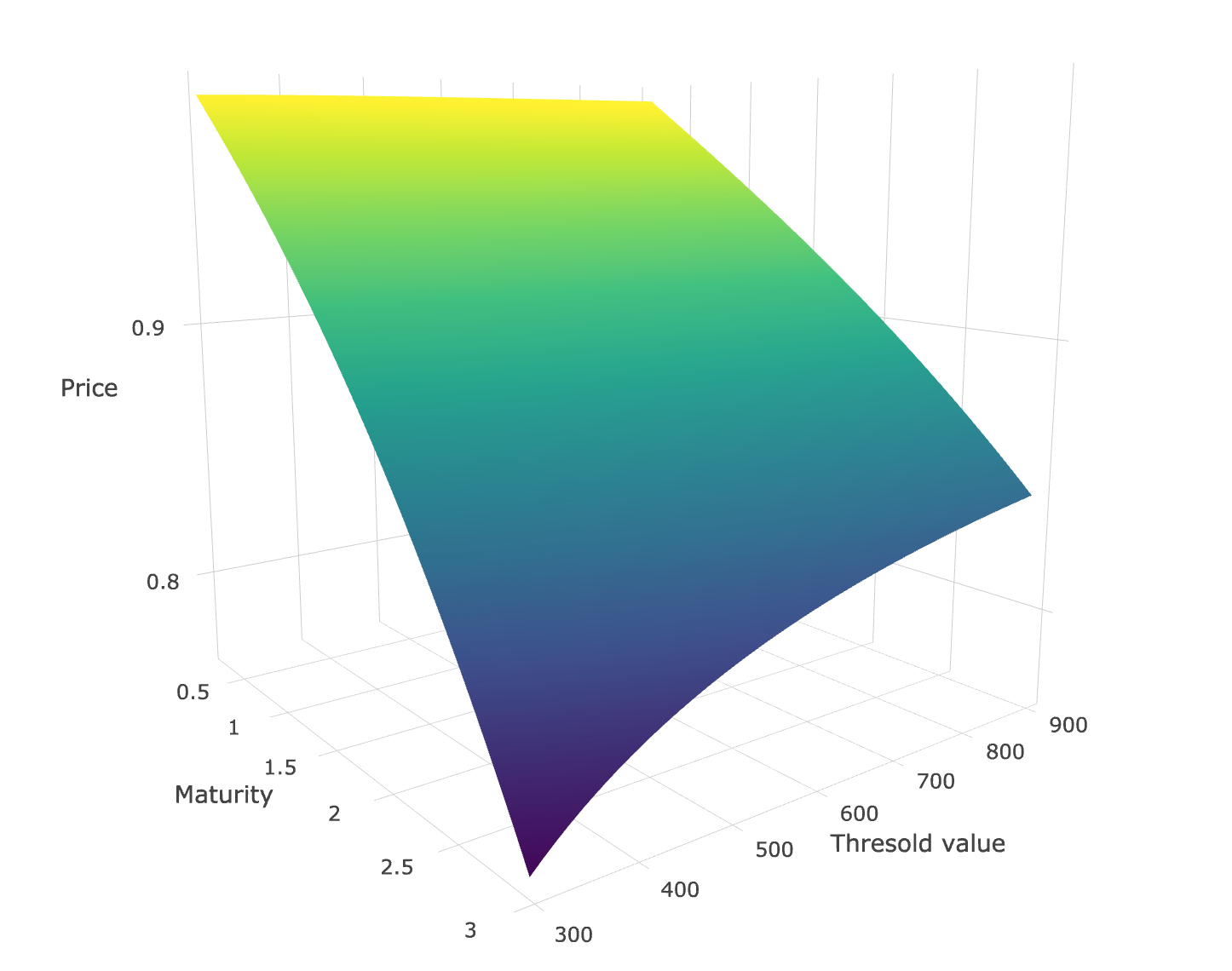}

\end{center}
\caption{The time $t=0$ surface of CAT bond price with respect to the threshold value and the maturity.}  \label{FigInt}
\end{figure}

\section{Conclusion}
In this paper, we have explored the abilities of two models related to credit risk modeling for CAT bond pricing. In Model 1, we have first investigated the case where the trigger event time is covered by totally inaccessible stopping times that make it to be totally inaccessible. We then study the second case where those stopping times are predictable hence the trigger event time is also predictable.  We have derived all the quantities of interest for pricing zero-coupon CAT bonds. We also showed that, in some settings, the model generalizes some ones developed in the CAT bond modeling.\\  
In Model 2 we started with the survival process of an investor in a CAT bond and then based on the results of enlargement of filtration, we construct the trigger event time. We have shown that this framework is related to the generalized Cox model \cite{gueye2021generalized} introduced in credit risk modeling and can lead to some closed-form prices of zero-coupon CAT bonds. We have studied the case where the aggregate loss process is a Shot-Noise. In that setting, the prices of zero-coupon CAT bonds admit some negative jumps at the occurrence times of the catastrophe events. Furthermore, we have illustrated these jumps in an example where we especially use a Markovian Shot-Noise which is more tractable for simulation. A comparison with a framework using the compound Poisson process (CPP) reveals that it is more suitable to use the Shot-Noise process in our approach for catastrophe loss modeling than using the CPP.\\

While the two approaches offer a new perspective on modeling CAT bonds, some potential improvements have to be taken into account in the forthcoming studies.\\ 
In model 1, we have shown that the zero-coupon CAT bond prices depend on the distribution of the aggregate losses. However, the estimation of this distribution could lead to some problems because of the low available data in the CAT bond framework. An adapted numerical approximation could be achieved for calibrating this model in real data.\\
In model 2, we did not interested in the estimation of the Shot-Noise since several approaches to estimating that process exists in the literature (see, e.g., \cite{Schmidt2014catastrophe} and the literature therein). However, it would be better to investigate how to incorporate the spatial resolution for hazard and exposure models developed in \cite{mistry2022pricing} in that estimation.

%%%%%%%%%%%%%%%%%%%%%APPENDIX

\appendix

\section*{Appendix}

\addtocounter{figure}{-6} 
\begin{figure}[hbt!]
\centering
\includegraphics[width=7cm, height = 5 cm] {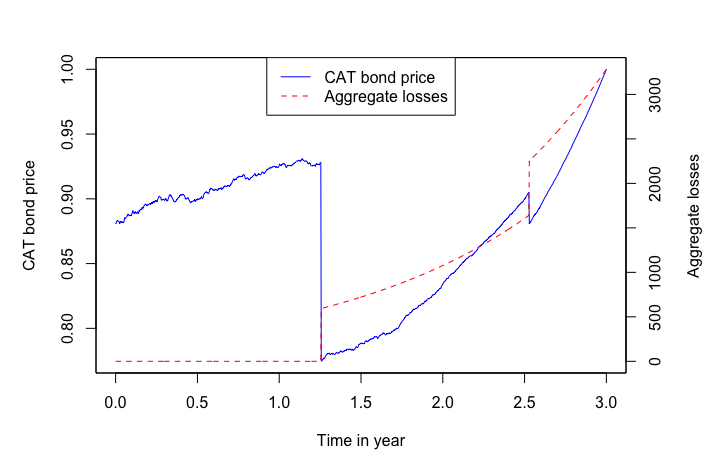}
\includegraphics[width=7cm, height = 5 cm] {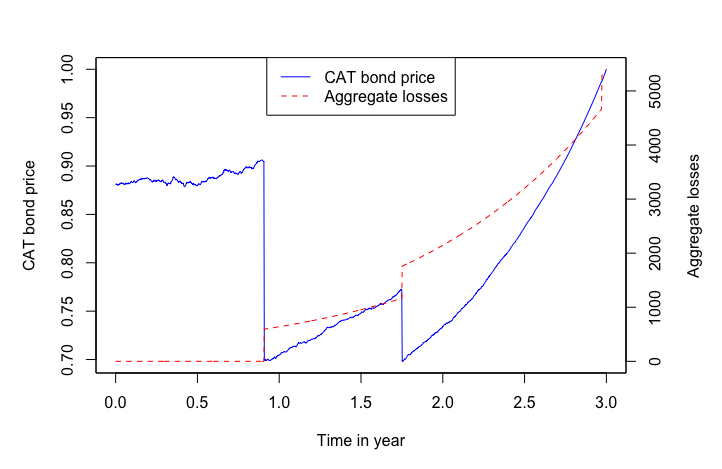}\\
\includegraphics[width=7cm, height = 5 cm] {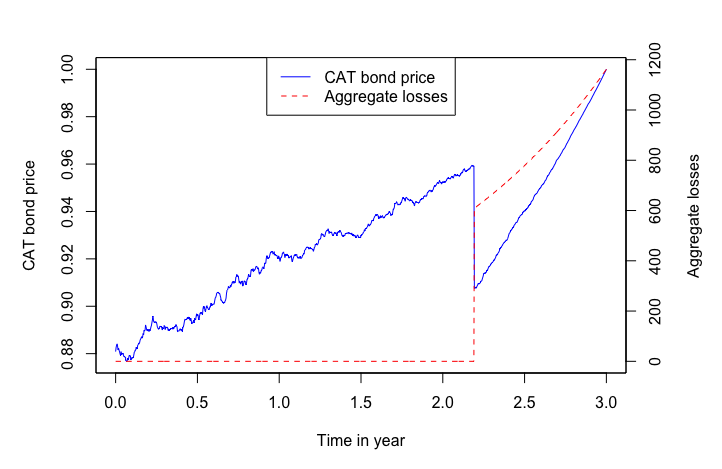}
\includegraphics[width=7cm, height = 5 cm] {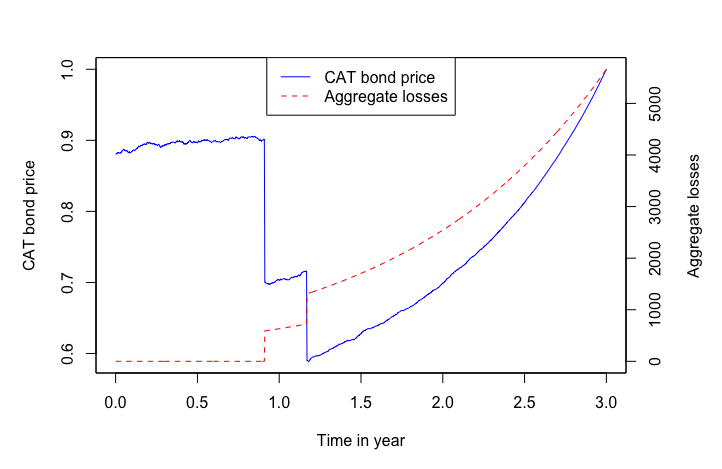}
\includegraphics[width=7cm, height = 5 cm] {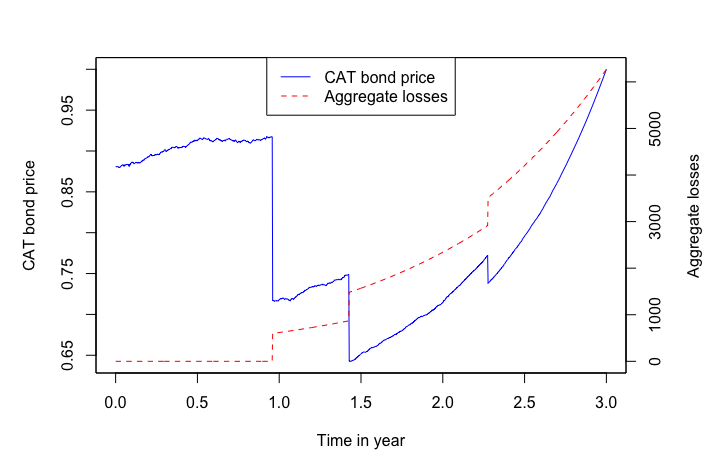}\includegraphics[width=7cm, height = 5 cm] {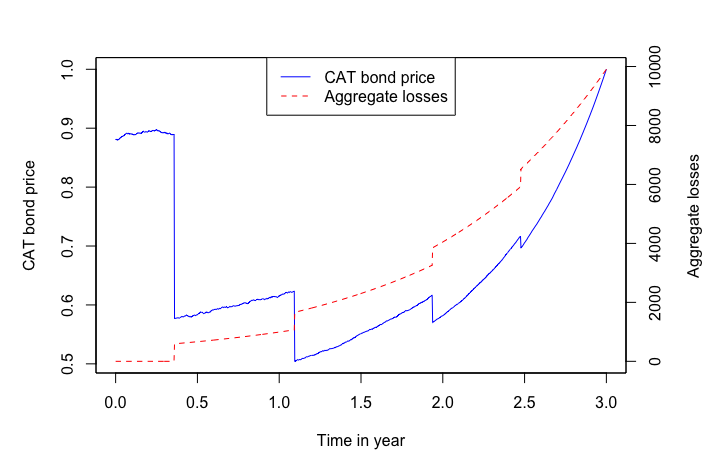}
\caption{Different scenarios for the CAT bond price vs Aggregate losses.}
 \label{Sc1}
\end{figure}

\newpage

\addtocounter{figure}{1}

\begin{figure}[hbt!]
\centering
\includegraphics[width=7cm, height = 5 cm] {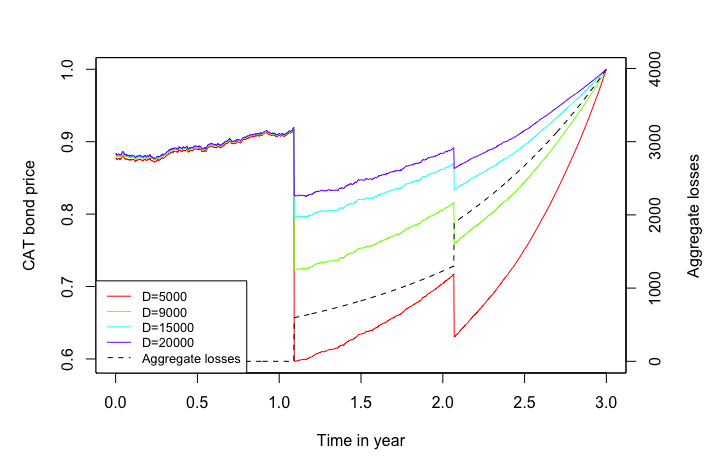}
\includegraphics[width=7cm, height = 5 cm] {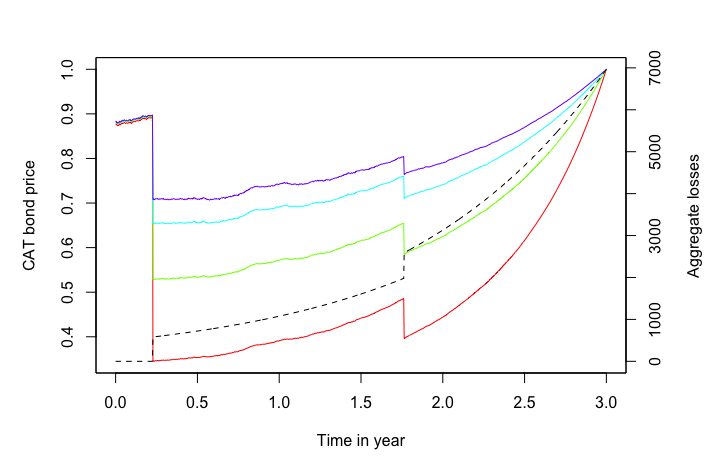}\\
\includegraphics[width=7cm, height = 5 cm] {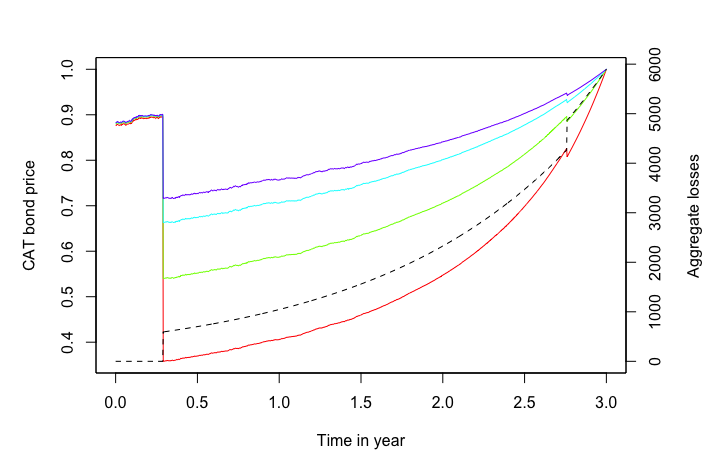}
\includegraphics[width=7cm, height = 5 cm]{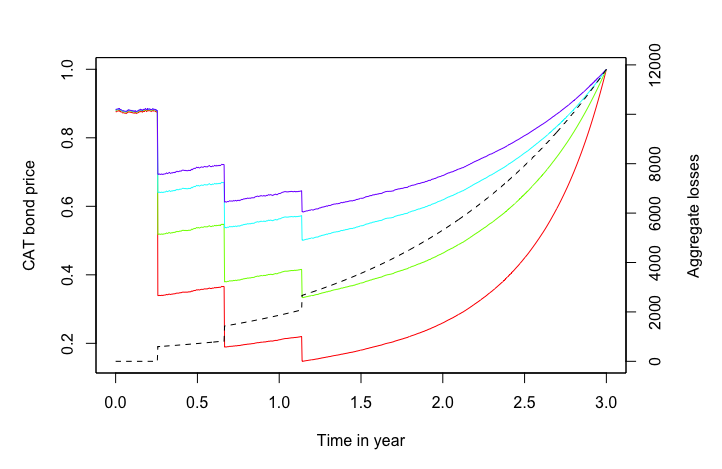}
\includegraphics[width=7cm, height = 5 cm] {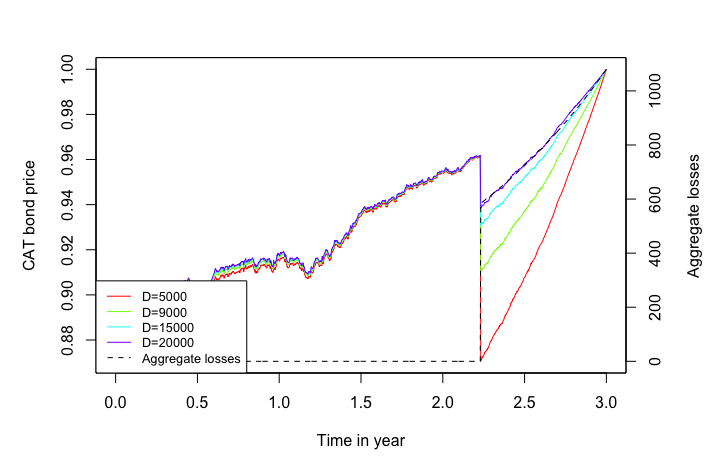}
\includegraphics[width=7cm, height = 5 cm] {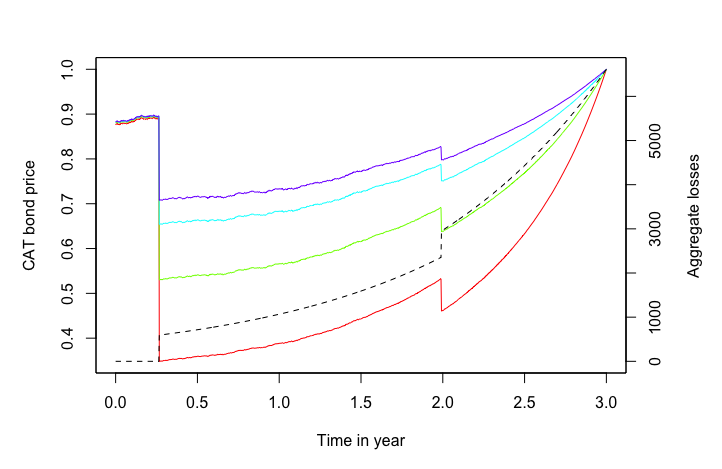}

\caption{Different scenarios for the CAT bond price with respect to different threshold values.}
 \label{Sc2}
\end{figure}

\addtocounter{figure}{1}

\begin{figure}[hbt!]
\centering
\includegraphics[width=7cm, height = 5 cm] {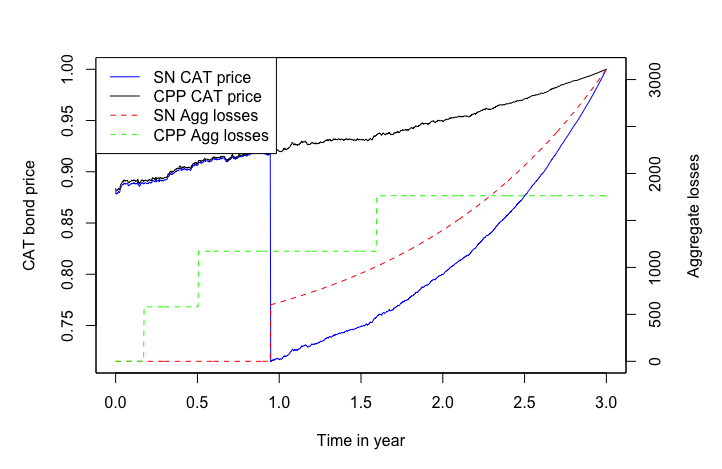}
\includegraphics[width=7cm, height = 5 cm] {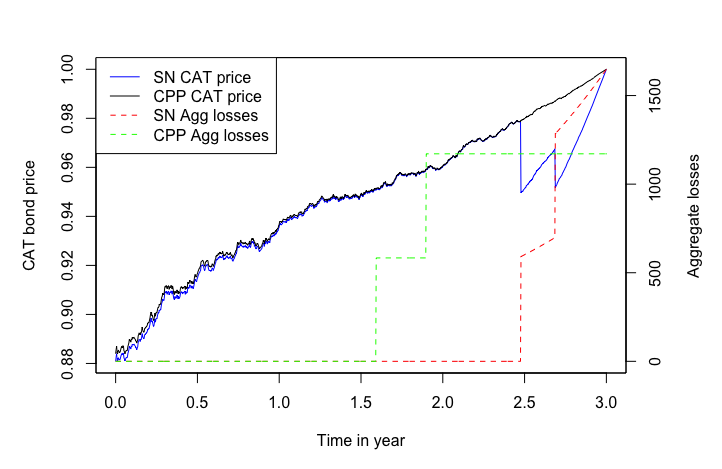}\\
\includegraphics[width=7cm, height = 5 cm] {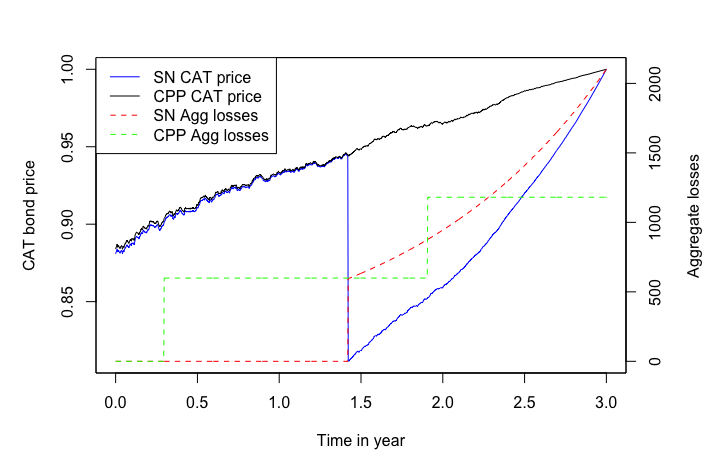}
\includegraphics[width=7cm, height = 5 cm] {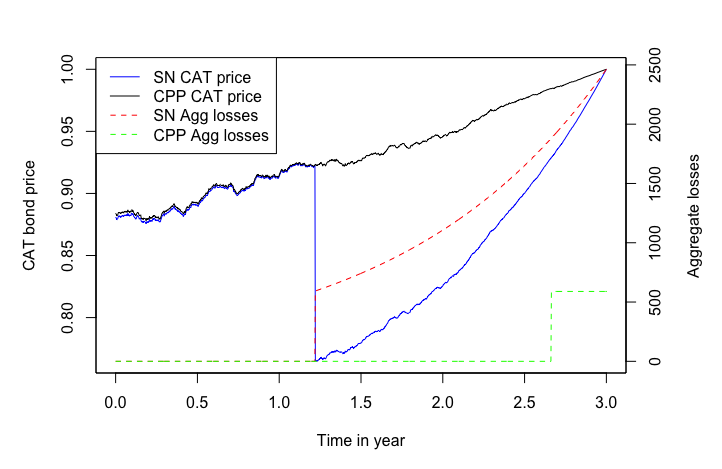}
\includegraphics[width=7cm, height = 5 cm] {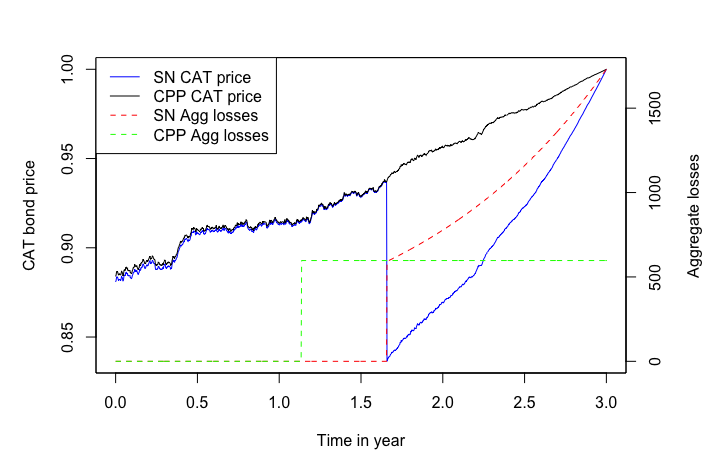}
\includegraphics[width=7cm, height = 5 cm] {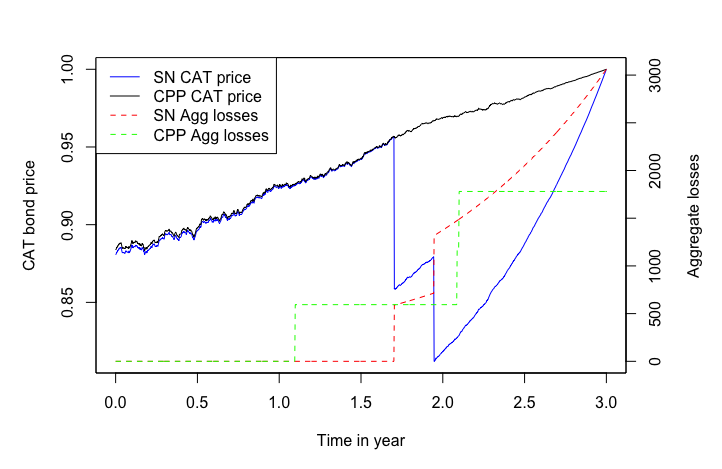}

\caption{Different scenarios for the CAT bond price via the Shot-Noise vs the CAT bond price using the CPP.}
 \label{Sc3}
\end{figure}

\end{document}